\documentstyle[preprint,aps]{revtex}
\def\be{\begin{equation}}
\def\ee{\end{equation}}
\def\bea{\begin{eqnarray}}
\def\eea{\end{eqnarray}}
\def\nn{\nonumber}
\def\th{\theta}
\def\ph{\phi}
\def\lt{\left}
\def\rt{\right}
\begin{document}
 \draft
\title{
Physical interpretation of constants in the solutions
to  the Brans-Dicke equations
}
\author{Aroonkumar Beesham\footnotemark[1]
\footnotetext[1]
{Department of Applied Mathematics, University of Zululand,Private Bag X1001,
Kwa-Dlangezwa 3886, South Africa.E-mail : abeesham@pan.uzulu.ac.za}
}
\maketitle
 \begin{abstract} 
Using an energy-momentum complex we give a physical interpretation
to the constants in the well-known static spherically symmetric asymptotically
flat vacuum solution to the Brans-Dicke equations.  The positivity of the tensor
mass puts a bound on  parameters in the solution.
 \end{abstract}

\section{Introduction}
It is well-known that scalar fields have been conjectured to give rise to long-range gravitational fields\cite{conj}.
Several theories involving scalar fields are known (see \cite{BD61}-
\cite{theories} and references therein).
Brans-Dicke (BD) gravity\cite{BD61} involves a scalar field and is perhaps the 
most viable alternative 
theory  to Einstein's general theory of  relativity.

This theory  is the simplest of the scalar-tensor theories and is a  
modification of  general relativity which accommodates   
Mach's principle as well as Dirac's large numbers hypothesis.
BD theory contains a massless scalar field $\Phi$ and a dimensionless
constant $\omega$ that describes the strength of the coupling between $\Phi$
and  the matter. It is usually believed that the post-Newtonian expansions of 
BD theory and general relativity agree in the limit $|\omega| \rightarrow \infty$.
Whilst this is usually the case, Romero and Barros\cite{RB93} worked out a 
number of examples when
this theory does not always go over to general relativity as 
$|\omega| \rightarrow \infty$. Recently, Banerjee
and Sen\cite{BS97} showed that BD theory goes over to general relativity
in the large $\omega$ limit only if the energy-momentum tensor of the source
term in the BD field equations is traceless.
This theory has satisfied all  the known  solar system experimental tests 
for $|\omega| \geq 500$\cite{Wil93}. Though the theory allows both positive as well as
negative values of $\omega$, it is usually assumed that $\omega > 0$ for
certain physical reasons. 

Though BD theory as well as the general theory of
relativity ``agree'' in the post-Newtonian limit described above, it is 
imperative to study
strong field cases in which the two theories could give different results.
These studies could provide experimental tests that might support one of them
and reject the other. 
One such avenue is gravitational waves (as opposed to 
general relativity theory, BD theory allows monopole as well as 
dipole radiation). The nature of black holes forming due to  gravitational
collapse could be another avenue to test these theories. The Hawking 
theorem\cite{Haw72} states that the Schwarzschild metric is the only
spherically symmetric solution of the vacuum Brans-Dicke field equations.
However, his theorem assumes the {\em weak energy condition} and considers
the scalar field $\Phi$ to be constant outside the black hole. 


Shortly after BD theory was proposed, one of the authors 
(Brans\cite{Bra62}) obtained the exact static vacuum solution to the BD
equations. The Brans type I solution is the only one which is permitted for
all values of $\omega$ (the other three forms are allowed only for
$\omega \leq -3/2$). Recently Campanelli and Lousto\cite{CL93} studied this
solution and found that for a certain range of parameters, the Brans type I
solutions represent black holes. It is possible  that a   realistic
gravitational collapse in BD theory could lead to the formation of ``Brans
black holes''.  However, not much (if any) studies have been done
to give physical interpretations to the constant parameters appearing
in the Brans solutions.  Using an energy-momentum complex\cite{Lee74}
in BD theory we investigate this problem. This is the aim of this paper.
We use  units in which the speed of light in vacuum $c$ and $\Phi_{\infty}$
(the value of the scalar field far from all sources) are unity.
The metric has signature $+ - - -$, and Latin (Greek)  indices take values 
$0\ldots3$  ($1\ldots3$).

\section{Scalar and tensor  mass  in BD theory}
The subject of energy-momentum localization has been debated since the beginning
of general relativity. Bergqvist\cite{Ber92} investigated seven different  
coordinate-independent definitions of quasi-local mass and found that no 
two of  them give the same result for the Reissner-Nordstr\"om (RN) and Kerr 
spacetimes.  Virbhadra and his collaborators (Rosen, 
Aguirregabiria and Chamorro) (\cite{Vir}-\cite{Colls}) showed that several energy-mometum
complexes coincide and give the same result (local values) for several
solutions in the Einstein as well as the Einstein-Maxwell theories when
calculations are carried out in ``cartesian coordinates''.
For any Kerr-Schild class solution, all the well-known energy-momentum complexes
 coincide\cite{Colls}.  Though this is
an encouraging result,  the long-standing problem cannot be considered to be
settled and  is still debatable. However, the total energy and momentum
of  asymptotically flat spacetimes are unambiguously accepted.  The situation
in BD theory is not better. The use of the energy-momentum complex in BD theory\cite{Lee74}
is also restricted to ``cartesian coordinates'', and only the total value
(integrated over all space) is unambiguously considered.
In BD theory, orbiting test particles (far away from a bounded system)
measure the total active gravitational mass (Keplerian mass) $M$ while orbiting test black
holes measure the tensor mass $M_T$\cite{Haw72}. The difference between the two
$M-M_T$ is the scalar mass. The tensor mass is always positive definite.
It decreases monotonically by emission of gravitational waves\cite{Lee74}.
The tensor mass is the active gravitational mass measured by a test Schwarzschild
black hole in the asymptotic region.

The tensor, scalar and the Keplerian masses are given by\cite{Lee74}
 \begin{equation}
M_T = \frac{1}{16\pi} \int \lt[\Phi^2 \Theta^{0 \alpha 0 \beta}\rt]_{,\alpha}
      d^2\Sigma_{\beta} ,
 \end{equation}

 \begin{equation}
M_S = \frac{1}{16\pi} \int \lt[\lt(\Phi^2-1\rt) \Theta^{0 \alpha 0 \beta}\rt]_{,\alpha}
      d^2\Sigma_{\beta} ,
 \end{equation}
and
 \begin{equation}
M = \frac{1}{16\pi} \int \lt[\lt(2 \Phi^2-1\rt) \Theta^{0 \alpha 0 \beta}\rt]_{,\alpha}
      d^2\Sigma_{\beta} ,
 \end{equation}
respectively. The quantity $\Theta^{mjnk}$ in the above expressions is given by
\begin{equation}
\Theta^{mjnk} = - g \lt(g^{mn} g^{jk} - g^{mk} g^{jn}\rt),
\end{equation}
which has the symmetries of the Riemann curvature tensor.
\section{The Brans solution}
The Brans-Dicke vacuum field equations are
\bea
R_{ik} &=& \frac{\omega}{\Phi^2} \Phi_{,i} \Phi_{,k} + \frac{\Phi_{;ik}}{\Phi}, \nn\\
\Box \Phi &=& 0.
\eea
There exist exact solutions to the above equations\cite{CL93}-\cite{Pim97}.
Static spherically symmetric and asymptotically flat exact solutions to the
above equations were given by Brans (see in \cite{CL93}), which are
expressed by the  line element
\be
ds^2 = A\lt(r\rt)^{m+1} dt^2 - A\lt(r\rt)^{n-1} dr^2 
      - A\lt(r\rt)^n \  r^2 \lt(d\th^2 + \sin^2\th d\phi^2\rt),
\ee
and the scalar field
\be
\Phi\lt(r\rt) =  A\lt(r\rt)^{- \frac{m+n}{2}},
\ee
where
\be
A\lt(r\rt) = 1 - 2 \frac{r_0}{r}.
\ee
The quantities $m,n,\Phi_0$ and $r_0$ are arbitrary constants. 

Putting $m = 0, n = 0$ in the above solutions gives the Schwarschild metric.
Campanelli and Lousto\cite{CL93} have demonstrated that the above solution
represents black holes if $n \leq -1$.

As it  is well-known that the use of  the energy-momentum complex is restricted
to quasi-cartesian coordinates (see\cite{Vir}-\cite{Colls} and references
therein), one transforms the  line element $(6)$ to these
coordinates, according to
\bea
x\ &=&\ r\ \sin\th\ \cos\ph, \nn\\
y\ &=&\ r\ \sin\th\ \sin\ph, \nn\\
z\ &=&\ r\ \cos\th,
\eea
and gets 
\be
ds^2 = A\lt(r\rt)^{m+1} dt^2 - A\lt(r\rt)^n \lt(dx^2+dy^2+dz^2\rt)
      - \frac{A\lt(r\rt)^{n-1}-A\lt(r\rt)^n}{r^2} \ \lt(x dx + y dy +z dz\rt)^2.
\ee

\section{Calculations}
As the metric under consideration is static, the momentum components will
vanish. Therefore, we evaluate only the masses associated with this
spacetime.  Using the line element given by Eq. $(10)$, we calculate the determinant and the 
contravariant  components of the metric tensor. 
Further we calculate the required components of $\Theta^{mjnk}$, which are

\bea
\Theta^{0101}&=& \frac{\lt(1-\frac{2r_0}{r}\rt)^{2n-1}}{r^3} 
                  \lt(-r^3+2 r_0 x^2\rt),        \nn\\
\Theta^{0202}&=& \frac{\lt(1-\frac{2r_0}{r}\rt)^{2n-1}}{r^3} 
                   \lt(-r^3+2 r_0 y^2\rt),           \nn\\
\Theta^{0303}&=& \frac{\lt(1-\frac{2r_0}{r}\rt)^{2n-1}}{r^3} 
                    \lt(-r^3+2 r_0 z^2\rt),          \nn\\
\Theta^{0102}&=&  \frac{2 \lt(1-\frac{2r_0}{r}\rt)^{2n-1}}{r^3} r_0 x y,\nn\\
\Theta^{0203}&=&  \frac{2 \lt(1-\frac{2r_0}{r}\rt)^{2n-1}}{r^3} r_0 y z,\nn\\
\Theta^{0301}&=&  \frac{2 \lt(1-\frac{2r_0}{r}\rt)^{2n-1}}{r^3} r_0 z x.
\eea 
Substituting the above in Eqs. $(1-3)$ and taking the limit 
$r \rightarrow \infty$ we get 
\bea
M_T &=& \frac{r_0}{2} \lt(m-n+2\rt), \nn\\
M_S &=& \frac{r_0}{2}   \lt(m+n\rt), \nn\\
M &=& r_0 \lt(m+1\rt).
\eea
It is clear that $M=M_T+M_S$.
Using Eqs. $(6)$, $(7)$, $(8)$ and $(12)$, the values of $g_{00}, g_{11}$
and $\Phi$ in the asymptotic region are
\bea
g_{00}&=& 1-\frac{2M}{r}, \nn\\
g_{11}&=& -1 + \frac{2}{r} \lt(M_S - M_T\rt), \nn\\
\Phi  &=& 1 + \frac{2M_S}{r}.
\eea
For the Schwarzschild metric ($m=0$ and $n=0$) Eq. $(12)$ gives
\bea
M_T &=& M = r_o , \nn\\
M_S &=& 0 ,
\eea
as expected. Now Eq. $(12)$ can be expressed as
\bea
r_o &=& M_{sch}, \nn\\
m   &=& \frac{M_S+M_T}{M_{sch}} - 1, \nn\\
n   &=& \frac{M_S-M_T}{M_{sch}} + 1,
\eea
where $M_{sch}$ stands for the Schwarzschild mass.
 To have the total mass of a Schwarzschild
black hole  positive, one considers $r_0 > 0$. Further, to respect the
positivity of the tensor mass associated with the Brans metric considered here, 
one has to put a restriction on the parameters $m$ and $n$ 
(i.e. $m-n+2 \geq 0$). 
\section{Discussion}
BD theory is the most viable alternative theory to Einstein's general
theory of relativity and it incorporates Mach's principle as well as 
Dirac's large numbers hypothesis. This theory has been supported by 
observational tests.
In recent years there has been renewed interest in BD theory as its 
application to the cosmological models of the universe (during the inflationary
era) makes bubble percolation more natural, and the low-energy  limit to
the theory of fundamental strings reduces to Brans-Dicke theory.
The Oppenheimer-Snyder collapse in BD theory leading to the formation
of black holes  suggested the possibility of black holes in this theory.
 With the help of an energy-momentum complex we have  obtained the tensor, 
scalar and  Keplerian masses associated with the Brans spacetime. 
 Though the spacetime is not flat for $m = -n = -1$, the scalar as well
as the tensor masses are zero.
Campanelli and Lousto\cite{CL93} showed that the Brans type I solutions 
represent
black holes if $n \leq 0$, but their investigations  do not put any restriction on the parameter
$m$. However, we have shown that the positivity of tensor mass puts a bound
$m \geq n-2$. Thus, one gets an interesting result that a physically realistic 
Brans black hole must be given by $n \leq 0$ with  $m \geq n-2$.
The physical interpretation of the constant parameters $r_o$, $m$ and $n$
in the Brans solution are clear from the Eq. $(14)$, i.e., these can be
expressed in terms the Schwarzschild, scalar and tensor masses.

\acknowledgments
Thanks are due to the FRD, South Africa, and to K S Virbhadra for 
enlightening discussions.

\end{document}